\RequirePackage{fix-cm}
\documentclass[twocolumn,epjc3]{svjour3}  
\smartqed  
\RequirePackage{graphicx}
\usepackage{amsmath}

\usepackage[backend=biber,style=numeric,sorting=none]{biblatex}
\RequirePackage{mathptmx}      
\RequirePackage{latexsym}

\RequirePackage[colorlinks,citecolor=blue,urlcolor=blue,linkcolor=blue]{hyperref}

\journalname{Eur. Phys. J. C}

\begin{document}

\title{Characterizing the Neutron Skin of $^{48}$Ca Through Collective Flow at the CERN Large Hadron Collider
}

\titlerunning{Study neutron skin at the LHC}        

\author{Andreas Vitsos\thanksref{addr1}
        \and
        Leonora Misciattelli Mocenigo Soranzo\thanksref{addr1} 
        \and You Zhou\thanksref{e1, addr1}
}

\thankstext{e1}{e-mail: you.zhou@cern.ch}


\institute{Niels Bohr Institute, Jagtvej 155A, 2200 Copenhagen, Denmark \label{addr1}
}

\date{Received: \today / Accepted: \today}

\maketitle

\begin{abstract}

The recently developed ``imaging-by-smashing’’ technique has emerged as a powerful approach to connect final-state collective flow phenomena in ultra-relativistic nuclear collisions with the intrinsic structure of the colliding nuclei. While most efforts have focused on constraining nuclear shape properties such as deformation and triaxiality, less attention has been given to the neutron skin, primarily in heavy nuclei such as $^{208}$Pb. In this work, a novel study of the neutron-skin thickness in $^{48}$Ca is presented, based on comparative analyses of $^{48}$Ca+$^{48}$Ca and $^{40}$Ca+$^{40}$Ca collisions at $\sqrt{s_{\rm NN}}=5.02$~TeV. Simulations within the AMPT framework are employed to investigate the impact of varying neutron-skin thicknesses $\Delta r_{np}$ for $^{48}$Ca on collective flow observables, including anisotropic flow coefficients and mean transverse momentum ($[p_{\rm T}]$) fluctuations. The triangular flow $v_3$, quadrangular flow $v_4$, as well as the variance and skewness of $[p_{\rm T}]$ fluctuations, display notable sensitivity to $\Delta r_{np}$. These findings indicate that calcium-isotope runs at the LHC could provide an independent and complementary approach to constraining $\Delta r_{np}$, with the potential to help resolve the current tension between PREX and CREX measurements.

\end{abstract}

\section{Introduction}\label{intro}

Ultra-relativistic nuclear collisions at the Large Hadron Collider (LHC) and the Relativistic Heavy Ion Collider (RHIC) provide a unique opportunity to investigate nuclear matter under extreme conditions~\cite{STAR:2005gfr,PHENIX:2004vcz,PHOBOS:2004zne,BRAHMS:2004adc,ALICE:2022wpn,CMS:2024krd}. The primary objective of these experiments is to create and characterise the Quark--Gluon Plasma (QGP), a state of matter in which quarks and gluons are deconfined~\cite{Lee:1978mf,Shuryak:1980tp}. A key feature of QGP formation is the presence of collective flow~\cite{ALICE:2011ab,ATLAS:2012at,CMS:2013wjq,ALICE:2016ccg,ALICE:2025luc}, representing the collective expansion of the final particles produced in the collisions. The systematic study of collective flow is crucial not only for extracting the precise transport properties of the QGP~\cite{Voloshin:2008dg, Heinz:2013th, Song:2017wtw}, but also for providing direct access to the initial geometry of the collision system and its event-by-event fluctuations~\cite{Gardim:2011xv,Niemi:2012aj,Gardim:2014tya,Li:2021nas}.
Given the close connections between final-state collective flow and the initial conditions, an emerging approach known as the ``imaging-by-smashing'' technique~\cite{Jia:2025wey} utilises collective flow in ultra-relativistic nuclear collisions to probe the structure of the colliding nuclei. Applications of this technique at both RHIC and the LHC have established a novel connection between high-energy collisions and traditional nuclear structure physics, which is typically performed at low energies. They provide insights into a variety of nuclear structure features concerning $^{76}$Zr, $^{76}$Ru,  $^{238}$U, $^{129}$Xe and  $^{208}$Pb~\cite{STAR:2024wgy, EMAILstar-publicationbnlgov:2025rot,ALICE:2018lao,ALICE:2021gxt,ALICE:2024nqd}, and the exploration of nuclear shape phase transitions at ultra-relativistic energies~\cite{Zhao:2024lpc}.  

Although nuclear structure studies in ultra‑relativistic nuclear collisions have been developed rapidly in the last few years (see a recent review in Ref.~\cite{Giacalone:2025vxa}), the study of \textit{neutron skin} remains primarily a subject of low‑energy nuclear physics, with investigations at high energy having been pursued only very briefly~\cite{Xu:2021vpn, Jia:2022qgl, Liu:2022kvz, Liu:2023pav, Giacalone:2023cet}. In neutron-rich nuclei, excess neutrons tend to accumulate along the outer surface of the nucleus, forming what is known as the \textit{neutron skin}. Its thickness, $\Delta r_{np}$, defined as the root mean square difference of the neutron and the proton distributions \cite{Reinhard:2010wz}, is strongly correlated with the density dependence of the symmetry energy in the nuclear Equation of State (EoS), and hence with the pressure of neutron-rich matter \cite{Chen:2010qx, Dieperink:2003vs, Lattimer:2023rpe}. 


This property has implications not only for finite nuclei but also for neutron star structure, including their mass–radius relation and tidal deformability~\cite{Alam:2016cli}, thereby exerting a profound impact on the associated astrophysical phenomena \cite{Tsang:2023vhh, Lasky:2013yaa, Raithel:2019ejc, Most:2021ktk}.
The recent PREX-II and CREX experiments have provided valuable constraints on $\Delta r_{np}$ for $^{208}$Pb and $^{48}$Ca through parity-violating electron scattering experiments~\cite{PREX:2021umo,CREX:2022kgg}. However, the extracted results are in tension, with PREX-II~\cite{PREX:2021umo} giving $\Delta r_{np} = 0.283 \pm 0.071$ fm for ${^{208}\text{Pb}}$ and CREX \cite{CREX:2022kgg} reporting $\Delta r_{np} = 0.121 \pm 0.05$ fm for $^{48}$Ca. 
The corresponding $L$-values are inconsistent, with $L_{PREX}= 76 \sim 165$ MeV and $L_{CREX}=0 \sim 51$ MeV \cite{Tagami:2022spb}. On the theoretical side, state-of-the-art \textit{ab initio} calculations predicted $\Delta r_{np} = 0.14-0.20$ fm~\cite{Hagen:2015yea, Hu:2021trw}, a value slightly lower than the PREX-II measurements. These discrepancies have given rise to significant discussions about the nuclear symmetry energy and the EoS of neutron-rich matter. In addition to the efforts at low energies, attempts have been made to probe the neutron skin using ultra-relativistic nuclear collisions as an alternative approach~\cite{Xu:2021vpn, Jia:2022qgl, Liu:2022kvz, Liu:2023pav, Giacalone:2023cet}. By applying global Bayesian fits to collective flow measurements in Pb–Pb collisions at the LHC within the hydrodynamic framework Trajectum~\cite{Giacalone:2023cet}, one gains quantitative access to nuclear density profiles in the initial conditions, thereby constraining the neutron skin thickness. The extracted value, $\Delta r_{np} = 0.217 \pm 0.058$ fm, is generally compatible with both PREX and \textit{ab initio} calculations, but does not resolve the existing tension.

Despite the developments mentioned above, the neutron skin of $^{48}$Ca has not yet been fully explored at high energies, and the existing investigations have primarily focused on the asymmetry between neutron and proton spectator nucleons~\cite{Zhang:2025voj}. To fully exploit the potential of the “imaging-by-smashing” technique, the present work investigates neutron-skin effects via collective flow in collisions between Ca nuclei at the LHC using the AMPT model~\cite{Lin:2004en}. In addition to the $^{48}$Ca+$^{48}$Ca collisions at $\sqrt{s_{\rm NN}} = 5.02$~TeV with varying $\Delta r_{np}$ for $^{48}$Ca, a comparative analysis of $^{40}$Ca+$^{40}$Ca collisions at the same energy is presented, providing a robust baseline for understanding size and density effects in the initial conditions. The anisotropic flow and radial flow, characterised by mean transverse momentum fluctuations of the produced final-state particles, are analysed to assess their sensitivity to different neutron-skin thicknesses of $^{48}$Ca. This study aims to identify which observables serve as suitable probes for constraining $\Delta r_{np}$ in ultra-relativistic nuclear collisions.

\section{Simulation setup and observables}
\label{Simulationsetup}
\subsection{AMPT Setup}

\begin{figure*}
\begin{center}
  \includegraphics[width=0.8\textwidth]{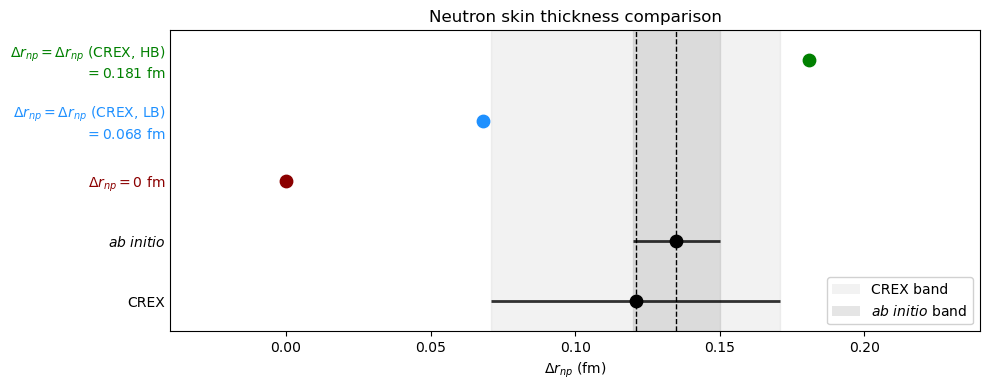}
\caption{\label{fig:1} Schema representing the values from CREX, ab initio and the parameters used in the simulations from Table \ref{tab:calcium_radii}.}  
\end{center}
\end{figure*}

The nuclear collisions between Ca nuclei at $\sqrt{s_{_{\rm NN}}}= 5.02$ TeV are simulated using the A MultiPhase Transport (AMPT) model \cite{Lin:2004en} with the string melting scenario, appropriate for systems where the formation of quark-gluon plasma (QGP) is expected. 
The dynamical evolution of the created system is modelled in four sequential stages. The initial conditions are generated using the Heavy Ion Jet Interaction Generator (HIJING) \cite{Wang:2000bf}, which provides the spatial and momentum distributions of incoming nucleons and produced mini-jets. Following HIJING, the parton interactions are modelled using Zhang's Parton Cascade (ZPC) Model \cite{Zhang:1997ej} with a partonic cross-section given by $\sigma = \frac{9\pi \alpha_s^2}{2 \mu^2}$, where $\alpha_s$ is the QCD coupling and $\mu$ is the screening mass. The hadronisation process is implemented via a quark coalescence mechanism \cite{Chen:2005mr} that recombines nearby partons into hadrons. This is essential for incorporating the string-melting scenario. The final-state hadronic rescattering is described by A Relativistic Transport (ART) model \cite{Li:1995pra}. Together, these components enable AMPT to capture the complete space-time evolution from the initial collision to the final state of the ultra-relativistic nuclear collisions.

In addition to the standard settings in the AMPT model, modifications have been implemented to investigate neutron-skin effects. Specifically, distinct Woods--Saxon distributions for neutrons and protons are employed to model the neutron-skin effect in the colliding nuclei, using the parameters that are illustrated in Fig.~\ref{fig:1} and listed in Table~\ref{tab:calcium_radii}. 
The proton and neutron distributions are assigned the same diffuseness parameter, $a_{0} = 0.586$~fm, for both $^{48}$Ca and $^{40}$Ca. To enable a meaningful comparison of neutron-skin effects, the proton radius $R_{p}$ for $^{48}$Ca is kept identical to that of $^{40}$Ca. In contrast, the neutron radius $R_{n}$ for $^{48}$Ca is varied across three scenarios: one with no neutron skin $\Delta r_{np}=0$; one with a small neutron skin, just outside the central range constrained by CREX $\Delta r_{np}=\Delta r_{np}$(CREX, LB); and one with a large neutron skin that exceeds the CREX range $\Delta r_{np}=\Delta r_{np}$(CREX, HB). 
All selected values are consistent with previous studies~\cite{Lattimer:2023rpe,Sammarruca:2023mxp,Schupp:2024owa}.

\begin{table*}
    \caption{The proton and neutron WS radii $R_p$, $R_n$, the diffuseness parameter $a$, the corresponding neutron skin values from the rms radii $\Delta r_{np}$ and number of simulated events for the Calcium isotopes. Reference values for the proton radius and diffuseness values are in range with values from \cite{Loizides:2014vua, Zenihiro:2018rmz}}
    \label{tab:calcium_radii}
    \centering
    \begin{tabular*}{\textwidth}{@{\extracolsep{\fill}}l l c c c c c@{}}
    \hline
    Case & Nucleus & \multicolumn{1}{c}{$R_p$ (fm)} & \multicolumn{1}{c}{$R_n$ (fm)} &
    \multicolumn{1}{c}{$a$ (fm)} & \multicolumn{1}{c}{$\Delta r_{np}$ (fm)} &
    \multicolumn{1}{c}{$N_{\text{Events}}$} \\
    \hline
     $^{40}$Ca & $^{40}$Ca & 3.385 & 3.375 & 0.586 & -- & $4.7\times10^{6}$ \\
    \hspace{0.5em} No neutron skin, $\Delta r_{np}=0$ & $^{48}$Ca & 3.387 & 3.387 & 0.586 & 0 & $4.1\times10^{6}$ \\
    \hspace{0.5em} CREX lower bound, $\Delta r_{np}$(CREX, LB) & $^{48}$Ca & 3.387 & 3.500 & 0.586 & 0.068 & $6.6\times10^{6}$ \\
    \hspace{0.5em} CREX higher bound, $\Delta r_{np}$(CREX, HB) & $^{48}$Ca & 3.387 & 3.685 & 0.586 & 0.181 & $4.8\times10^{6}$ \\
    \hline
    \end{tabular*}
\end{table*}

\subsection{Anisotropic flow}
In ultra-relativistic heavy-ion collisions, the produced particles exhibit collective behaviour manifested as anisotropies in their azimuthal distribution \cite{Ollitrault:1992bk, Snellings:2011sz, Voloshin:2007pc}. These anisotropies reflect the response of the strongly interacting medium to the initial-state geometry and fluctuations, and constitute a crucial observable for probing the collective dynamics of the collision process.
The azimuthal angle $\varphi$ distribution can be expressed in terms of Fourier coefficients \cite{Voloshin:1994mz}:  
\begin{equation} 
\hspace{0.5cm} E \frac{d^3N}{d^3p}  \propto  1+ \sum_{n=1}^{\infty} 2v_n \cos[n(\varphi-\Psi_n)],
\end{equation}  
where $\varphi$ is the azimuthal angle of the produced particles. The $\Psi_n$ is the orientation of the $n$th harmonic flow symmetry plane. The flow coefficients $v_n$ quantify the magnitude of the anisotropic flow, with elliptic flow $n=2$, triangular flow $n=3$, quadrangular flow $n=4$, and so on. Since the flow symmetry plane $\psi_n$ is not directly available in experiments, $v_n$ coefficients are usually extracted from azimuthal correlations among final-state particles \cite{Borghini:2000sa}. The two-particle correlations of harmonic $n$, $c_n\{2\}$, and the corresponding flow coefficient $v_n\{2\}$ can be obtained via
\begin{align} \label{definev2n} 
\hspace{0.5cm} c_n\{2\} &= \langle \langle 2\rangle \rangle = \langle \langle e^{in(\varphi_1 - \varphi_2)} \rangle \rangle = \langle v_n^2 \rangle + \delta_n, \\
 v_n\{2\} &= \sqrt{c_n\{2\}}.
\end{align}  
Here $\delta_n$ denotes the so-called non-flow contribution, arising from correlations unrelated to collective motion. Reliable extraction of $v_n\{2\}$, therefore, requires the suppression of non-flow, which can be achieved by applying the $\eta$-gap method~\cite{Zhou:2015iba}.
For this paper, a pseudorapidity gap of $|\Delta \eta| > 0.2$ is applied to suppress non-flow correlations while preserving statistical significance consistently across results.
The study of flow coefficients can also be performed using multi-particle cumulants, which, by definition, eliminate lower-order (few-particle) correlations and are therefore less sensitive to nonflow effects~\cite{Borghini:2000sa}. Moreover, multi-particle cumulants carry independent information about event-by-event flow fluctuations~\cite{Voloshin:2007pc,Zhou:2015eya,Moravcova:2020wnf}. 
However, analysing the multi-particle cumulants of $v_{n}$ requires substantial statistics, which are not accessible within the scope of the existing AMPT simulations used in this study.

\subsection{Mean transverse momentum fluctuations}
Another set of powerful observables, which utilise the information of the mean transverse momentum of produced particles, are the multi-particle $p_{\rm T}$ cumulants, $\kappa_n$ \cite{Broniowski:2009fm, Bozek:2012fw,Giacalone:2020lbm}. In this study, the first, second, and third-order cumulants are employed, defined using the event-by-event $p_\text{T}$ correlations, $[p_\text{T}^{(m)}]$.
\begin{equation}
   \hspace{0.5cm} [p_\text{T}^{(m)}] = \frac{\displaystyle\sum^{M}_{k_1\neq...\neq k_m} w_{k_1}\cdot \ldots \cdot w_{k_m}p_{\text{T},k_2} \cdot \ldots \cdot p_{\text{T},k_m}}{\displaystyle\sum^{M}_{k_1\neq...\neq k_m} w_{k_1}\cdot \ldots \cdot w_{k_m}}.
\end{equation}
Here, $M$ is the number of particles in the selected kinematic range, and $w_i$ are particle weights accounting for detector inefficiencies. 
Then the multi-particle $p_{\rm T}$ cumulants $\kappa_n$ can be obtained
\begin{align}
 \hspace{0.5cm}   \kappa_1 &= \langle [p_{\text{T}}^{(1)}] \rangle,    \label{eq:k1}\\
    \kappa_2 &= \langle [p_{\text{T}}^{(2)}] \rangle - \langle [p_{\text{T}}^{(1)}] \rangle^2, \label{eq:k2}\\
    \kappa_3 &= \langle [p_{\text{T}}^{(3)}] \rangle - 3\langle [p_{\text{T}}^{(2)}] \rangle \langle [p_{\text{T}}^{(1)}] \rangle + 2\langle [p_{\text{T}}^{(1)}] \rangle^3,
    \label{eq:k3}
\end{align}
where $\langle \cdot \rangle$ denotes the average across all events. These $\kappa_n$ cumulants are obtained using the {\it Generic Algorithm} of multi-particle $p_{\rm T}$ correlations~\cite{Nielsen:2023znu}, within the kinematic regions of $|\eta| < 0.8$ and $0.2 < p_\text{T} < 5.0$ GeV/$c$.

\section{Results and discussion}

In this section, the sensitivity of collective flow observables to neutron skin thickness is examined by analyzing $^{40}$Ca+$^{40}$Ca and $^{48}$Ca+$^{48}$Ca collisions at $\sqrt{s_{_{\rm NN}}} = 5.02$ TeV, simulated with the AMPT model. Three neutron skin scenarios for $^{48}$Ca are considered: no neutron skin ($\Delta r_{np} = 0$), the CREX lower bound value ($\Delta r_{np} = 0.068$ fm), and the CREX upper bound value ($\Delta r_{np} = 0.181$ fm). The section is organized as follows: analysis results of the observable $\mathcal{O}$ as a function of centrality are presented. For each figure, the top panel displays the observable $\mathcal{O}$ in $^{40}$Ca+$^{40}$Ca and $^{48}$Ca+$^{48}$Ca collisions, while the middle panel shows the ratio of each $^{48}$Ca case to the $^{40}$Ca case, defined as $R(\mathcal{O}) = \frac{\mathcal{O}{^{48}\text{Ca}}}{\mathcal{O}{^{40}\text{Ca}}}$. The bottom panel presents the ratio between different $^{48}$Ca configurations, defined as $\tilde{R}(\mathcal{O}) = \frac{\mathcal{O}{^{48}\text{Ca}(\alpha)}}{\mathcal{O}_{^{48}\text{Ca}(\beta)}}$, where $\alpha$ and $\beta$ represent different scenario of neutron skins of $^{48}$Ca. The $R(\mathcal{O})$ and $\tilde{R}(\mathcal{O})$ ratios are of particular interest, as they help cancel the final-state effects arising from the system's dynamic evolution, which allows probing the size/density effects and the neutron-skin-related effects, respectively. A similar approach is employed for the isobaric runs of $^{96}\text{Ru}$ and $^{96}\text{Zr}$ collisions at RHIC~\cite{Jia:2022qgl}, which has been used to investigate the impacts from nuclear structure effects. The purpose of these comparisons is to assess whether the effects associated with neutron skin thickness can be observed via the observables of anisotropic flow and mean transverse momentum fluctuations in ultra-relativistic nuclear collisions at the LHC.

\subsection{Flow Analysis}

\begin{figure}[h!]
\includegraphics[width = 0.45\textwidth]{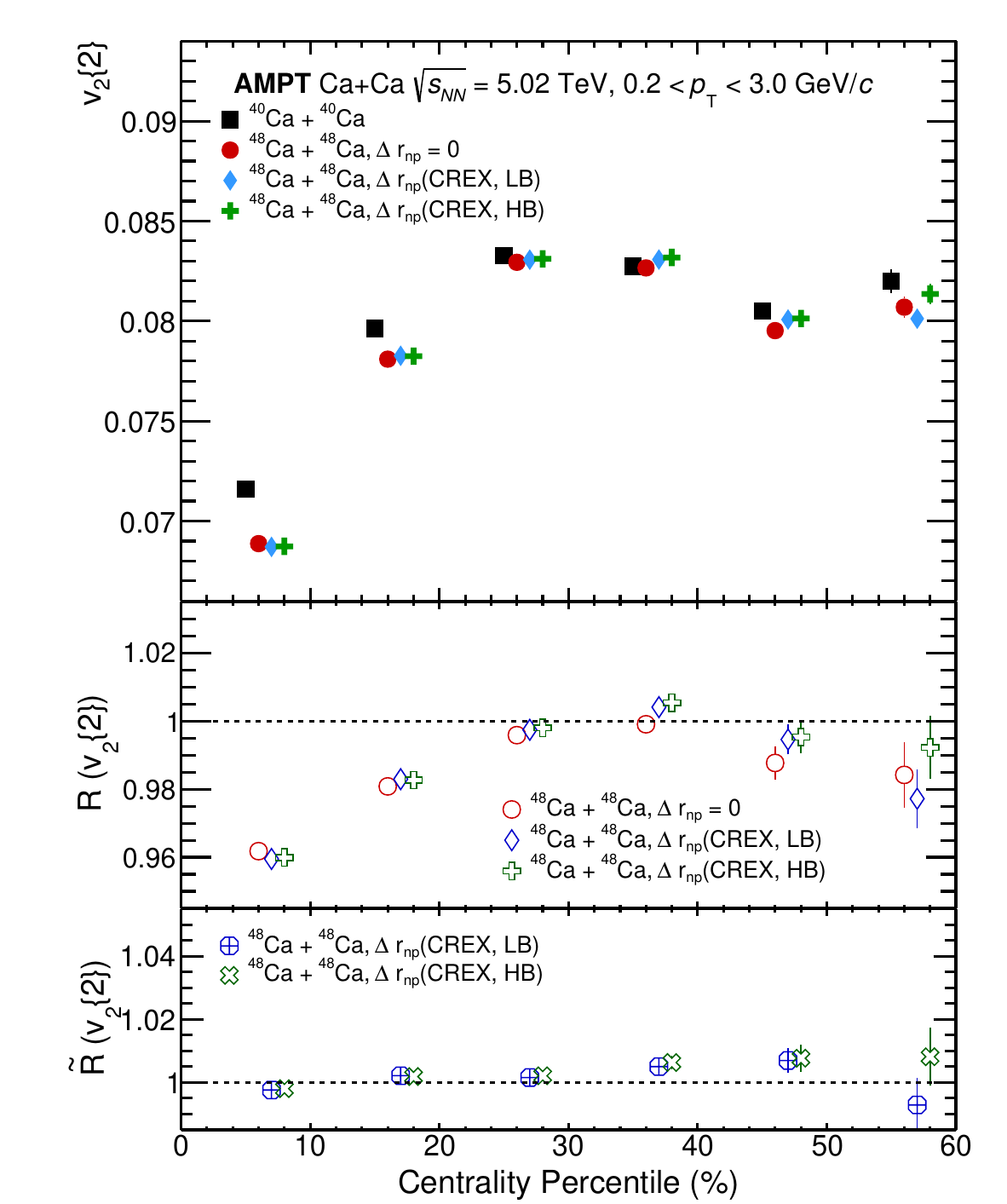}
\caption{The centrality dependence of $v_2\{2\}$ in Ca+Ca collisions (top), the ratios of each $^{48}$Ca+$^{48}$Ca case over the $^{40}$Ca+$^{40}$Ca case (middle) and the ratios with the $^{48}$Ca+$^{48}$Ca with $\Delta r_{np}=0$ (bottom).}
\label{v2}
\end{figure}

Figure \ref{v2} (top panel) shows the centrality dependence of $v_2\{2\}$ in $^{40}$Ca+$^{40}$Ca, and also in $^{48}$Ca+$^{48}$Ca collisions with the three different $\Delta r_{np}$. 
The $v_{2}\{2\}$ results from all cases exhibit a clear centrality dependence, reflecting the sensitivity of the final-state elliptic flow to the initial geometry of the overlap region between the two colliding Ca nuclei. This centrality dependence seems more significant when compared to the $v_{2}$ in the light-ion collisions~\cite{ALICE:2025luc} but less pronounced compared to those in Xe+Xe and Pb+Pb collisions at the LHC~\cite{ALICE:2011ab, ALICE:2016ccg, ALICE:2018lao, ALICE:2024nqd}. 
In the middle panel of Fig.~\ref{v2}, the ratios with respect to the $^{40}$Ca+$^{40}$Ca results allow a clear distinction between results from $^{40}$Ca+$^{40}$Ca and $^{48}$Ca+$^{48}$Ca collisions in central events (centrality range 0--20\%). The $v_{2}\{2\}$ values for the $^{48}$Ca+$^{48}$Ca system are approximately $4\%$ lower than those for $^{40}$Ca+$^{40}$Ca collisions. In particular, $v_{2}\{2\}$ in $^{48}$Ca+$^{48}$Ca collisions with $\Delta r_{np} > 0$ is smaller than in $^{40}$Ca+$^{40}$Ca collisions, despite the nuclei having nearly identical sizes. This difference can be attributed to the fact that fewer nucleons are packed within the same volume in  $^{48}$Ca as in the $^{40}$Ca nucleus, leading to more pronounced event-by-event fluctuations in the nucleon distribution. These enhanced initial fluctuations lead to stronger final-state flow fluctuations, ultimately increasing the $v_{2}\{2\}$ values. However, it is not possible to distinguish between the different neutron skin thicknesses only based on $v_2\{2\}$ presented here. This is highlighted in the bottom panel of Fig.~\ref{v2}, which shows the ratio between the $v_2\{2\}$ of each case with respect to the results in $^{48}$Ca+$^{48}$Ca with $\Delta r_{np}$(CREX, LB) case. 
As shown, the $v_2\{2\}$ results exhibit negligible sensitivity to the neutron skin thickness of $^{48}$Ca. Therefore, $v_2\{2\}$ might not be an ideal probe for neutron skin thickness in ultra-relativistic nuclear collisions with Ca nuclei.

\begin{figure}[h!]
\includegraphics[width = 0.45\textwidth]{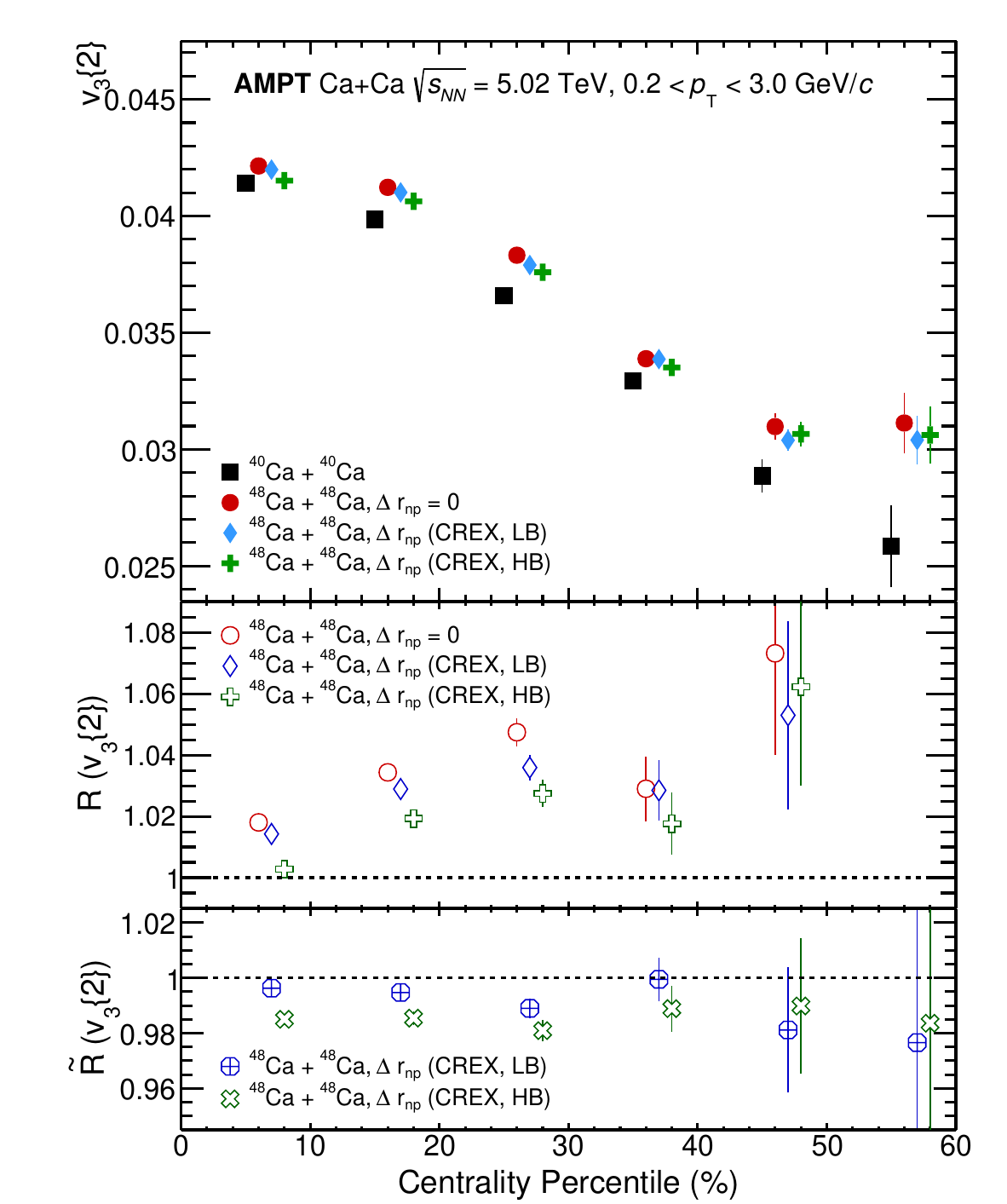} 
\caption{\label{v3} Centrality dependence of $v_3\{2\}$ in Ca+Ca collisions (top), the ratios of each $^{48}$Ca+$^{48}$Ca case over the $^{40}$Ca+$^{40}$Ca case (middle) and the ratios with the $^{48}$Ca+$^{48}$Ca with $\Delta r_{np}=0$ (bottom).}
\end{figure}

Besides the elliptic flow $v_2\{2\}$ study, the triangular flow $v_3\{2\}$ is primarily driven by initial triangularity, originating from the initial spatial fluctuations~\cite{ALICE:2011ab,Alver:2010gr}. It might be more sensitive to size fluctuations arising from different neutron-skin effects, compared to $v_2\{2\}$.
Figure~\ref{v3} presents the centrality dependence of $v_3\{2\}$ in $^{40}$Ca+$^{40}$Ca and $^{48}$Ca+$^{48}$Ca collisions. The $v_3\{2\}$ results decrease with increasing centrality percentile. This centrality dependence differs significantly from that observed in heavy-ion collisions~\cite{ALICE:2011ab,ALICE:2016ccg, ALICE:2018lao, ALICE:2024nqd}, yet it is consistent with recent findings in light-ion collisions at the LHC~\cite{ALICE:2025luc}. Moreover, the trend deviates from the centrality dependence of the initial triangularity $\varepsilon_3$~\cite{Loizides:2025ule}, indicating non-trivial dynamical evolution in the created system from ultra-relativistic collisions with Ca nuclei. For the centrality range 0--30\%, visible differences among the various $v_{3}\{2\}$ results are observed in Fig.~\ref{v3} (top panel), with the $^{48}$Ca+$^{48}$Ca collisions at $\Delta r_{np}=0$ yielding the largest values and the $^{40}$Ca+$^{40}$Ca collisions giving the smallest. Statistical uncertainties increase for more peripheral collisions, where definitive conclusions cannot be drawn with the current statistics.

These differences are better quantified in the ratio studies presented in the middle and bottom panels. For the $v_{3}\{2\}$ ratio of $^{48}$Ca+$^{48}$Ca with $\Delta r_{np} = 0$, and $^{40}$Ca+$^{40}$Ca, it deviates from unity by approximately 2\% in the central collision, which increases up to about 8\% in the peripheral collisions. This difference is explained by the fact that when more nucleons are packed in the same nucleus size, the higher nucleon density of $^{48}$Ca gives stronger density fluctuations, which in turn gives a larger $v_{3}\{2\}$ result. When the neutron skin thickness is further increased, the nucleon density decreases, leading to weaker density fluctuations and eventually a decrease in $v_{3}\{2\}$. The neutron skin effect is better accessed in the $\tilde{R}(v_{3}\{2\})$ ratio shown in the bottom panel, where the results from $\Delta r_{np}$(CREX, LB) show a minor difference w.r.t. the one from $\Delta r_{np}=0$. Meanwhile, the results from $\Delta r_{np}$(CREX, HB) exhibit a deviation from unity of 1.5--2\% for the presented centrality ranges. This reveals the imprint of neutron skin effects on the final-state triangular flow, which is more pronounced when the variation in neutron skin thickness is significant. Therefore, the $v_{3}\{2\}$ observable can serve as a useful probe for the neutron skin study of $^{48}$Ca in future ultra-relativistic collisions with Ca nuclei.

\begin{figure}[h!]
\includegraphics[width = 0.45\textwidth]{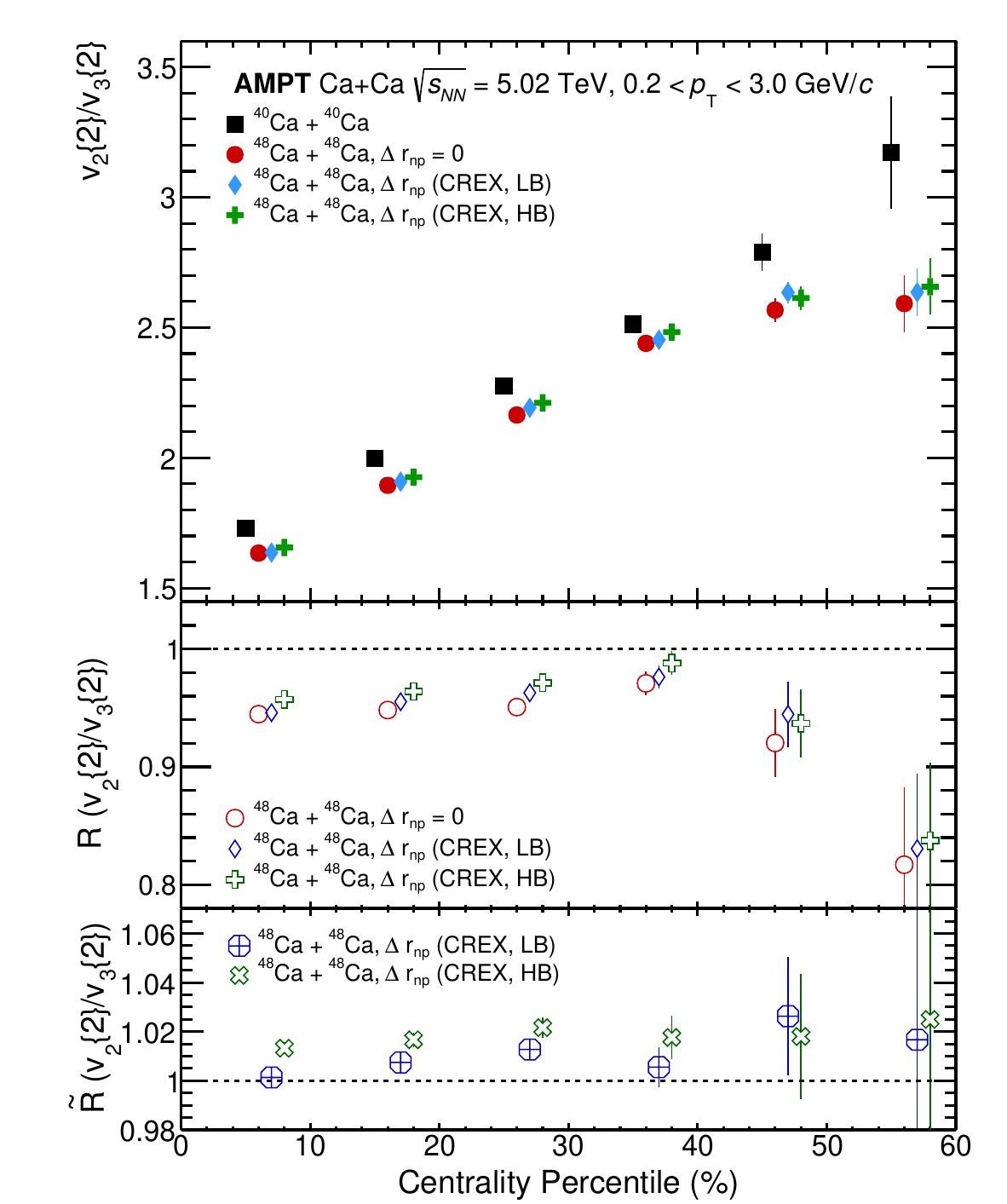}
\caption{\label{flowratio} Centrality dependence of $\frac{v_2\{2\}}{v_3\{2\}}$ in Ca+Ca collisions (top), the ratios of each $^{48}$Ca+$^{48}$Ca case over the $^{40}$Ca+$^{40}$Ca case (middle) and the ratios with the $^{48}$Ca+$^{48}$Ca with $\Delta r_{np}=0$ (bottom).}
\end{figure}

Both elliptic flow $v_2$ and triangular flow $v_3$ are driven by their corresponding initial spatial anisotropies, $\varepsilon_2$ and $\varepsilon_3$, and are influenced by the collective evolution of the system. The absolute magnitudes of $v_2\{2\}$ and $v_3\{2\}$ therefore reflect both geometry and dynamical effects. Their ratio $\frac{v_2\{2\}}{v_3\{2\}}$, however, can provide a more direct insight into the initial geometry of the system through an approximate cancellation of the responses to the evolution of the system, leaving a quantity that is closely correlated with the initial spatial anisotropies. For this reason, the ratio has been considered a potential probe to constrain neutron-skin effects in experimental studies.
The AMPT calculations of $\frac{v_2\{2\}}{v_3\{2\}}$ as a function of centrality are presented in Fig. \ref{flowratio}.
The results suggest that $\frac{v_2\{2\}}{v_3\{2\}}$ can effectively differentiate between the $^{40}$Ca+$^{40}$Ca and $^{48}$Ca+$^{48}$Ca collision systems. In the central collisions, the ratios differ by $\sim$ 5\% between the two collision systems (see Fig. \ref{flowratio}, middle panel). In addition, the sensitivity to the different neutron-skin thickness is investigated in the bottom panel of Fig. \ref{flowratio}. For central collisions, the ratio is enhanced by $\sim 1\%$ for $\Delta r_{np}=r_{\mathrm{CREX}}$, higher bound relative to $r_{\mathrm{CREX}}$, lower bound, while it displays a decreasing trend in the 0--30\% centrality range for the case with absence of neutron skin ($r_{np}=0$). This highlights the potential of $\frac{v_2\{2\}}{v_3\{2\}}$ as a potential probe of neutron skin thickness in ultra-relativistic nuclear collisions.

\begin{figure}[h!]
\includegraphics[width = 0.45\textwidth]{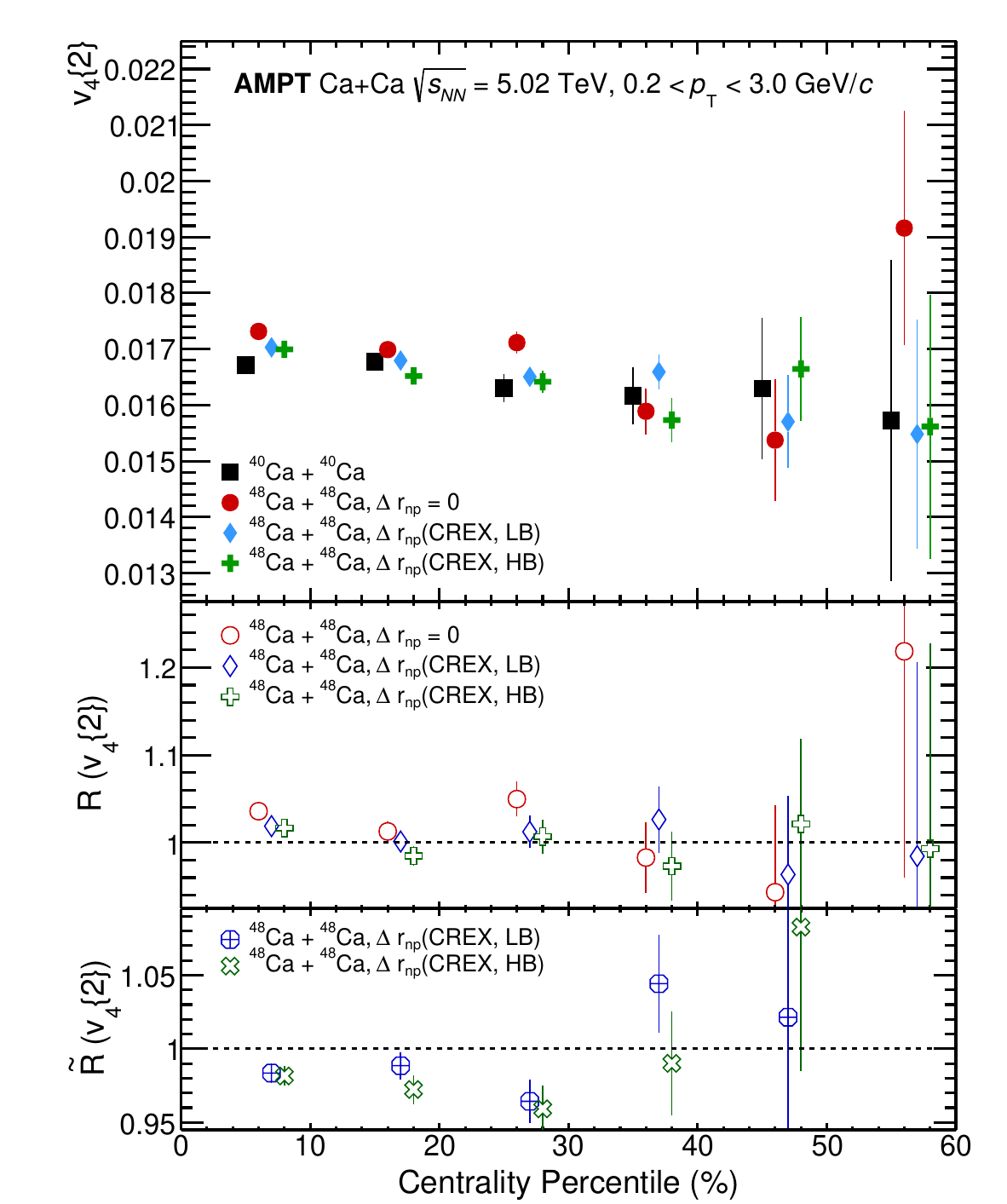}
\caption{\label{v4} Centrality dependence of $v_4\{2\}$ in Ca+Ca collisions (top), the ratios of each $^{48}$Ca+$^{48}$Ca case over the $^{40}$Ca+$^{40}$Ca case (middle) and the ratios with the $^{48}$Ca+$^{48}$Ca with $\Delta r_{np}=0$ (bottom).}
\end{figure}

Different from elliptic flow $v_{2}$ and triangular flow $v_{3}$, which exhibit an almost linear response to the initial $\varepsilon_{2}$ and $\varepsilon_{3}$, the quadrangular flow $v_{4}$ receives contributions not only from the linear response to the initial $\varepsilon_{4}$ but also from the non-linear term $\varepsilon_{2}^{2}$~\cite{ALICE:2017fcd,Yan:2015jma}. It is therefore of interest to explore how the final-state higher harmonic flow, such as $v_{4}$, reflects variations in neutron skin thickness. The centrality dependence of $v_{4}\{2\}$ in $^{40}$Ca+$^{40}$Ca collisions, as well as in $^{48}$Ca+$^{48}$Ca collisions with three different values of $\Delta r_{np}$, is presented in Fig.~\ref{v4}. The overall magnitudes of $v_{4}\{2\}$ are smaller compared to $v_{2}\{2\}$ and $v_{3}\{2\}$, and the statistical uncertainties become significant in peripheral collisions. A relatively weak centrality dependence is observed across all cases, with a slight decrease toward more peripheral collisions. The $v_{4}\{2\}$ values are largest for $^{48}$Ca+$^{48}$Ca collisions with $\Delta r_{np}=0$ in the 0--30\% centrality range, while the other cases are slightly smaller. This comparison is more evident in the ratio studies shown in the middle and bottom panels. The $v_{4}\{2\}$ results from $^{48}$Ca+$^{48}$Ca collisions with $\Delta r_{np}=0$ are approximately 2--3\% larger than those from $^{40}$Ca+$^{40}$Ca collisions in the 0--30\% centrality range. Since the linear contribution from $\varepsilon_{4}$ dominates in this centrality interval~\cite{ALICE:2017fcd, ALICE:2020sup}, the small difference in $v_{4}\{2\}$ can be attributed to the impact of nucleon density and its fluctuations on the initial $\varepsilon_{4}$, which subsequently affect the final-state $v_{4}\{2\}$. Meanwhile, the difference in $v_{4}\{2\}$ between $\Delta r_{np}$(CREX, LB) and $\Delta r_{np}=0$ is on average $\sim 2\%$ in central collisions, which is also the case when comparing the $v_{4}\{2\}$ results from $\Delta r_{np}$(CREX, HB) with that of $\Delta r_{np}=0$. These differences reflect the light mark of neutron skin effects on the final-state $v_{4}\{2\}$, showing the potential of $v_{4}\{2\}$ as a probe to constrain the neutron skin thickness of $^{48}$Ca.

A future experimental comparison of $v_3\{2\}$, $v_4\{2\}$ and $\frac{v_2\{2\}}{v_3\{2\}}$ ratio in $^{40}$Ca+$^{40}$Ca and $^{48}$Ca+$^{48}$Ca collisions at the LHC has the potential to shed light on the role of neutron skin thickness. Given that the observed differences in the AMPT studies are on the order of 1--2\%, it may be challenging to detect them in experiments, considering the current systematic uncertainties of $v_{n}\{2\}$ measurements~\cite{ALICE:2024nqd}. Thus, conducting isotope runs of $^{40}$Ca and $^{48}$Ca under matched experimental conditions, such as identical beam energies, detector configurations, and analysis procedures, will be crucial. It would largely cancel out the final systematic uncertainties in flow measurements, as done in the iosbar runs at RHIC~\cite{EMAILstar-publicationbnlgov:2025rot}, or light-ion runs at the LHC~\cite{ALICE:2025luc}. This setup enhances the possibility of detecting effects arising from differences in the neutron skin. 

\subsection{Mean transverse momentum fluctuations analysis}

\begin{figure}[h]
\includegraphics[width = 0.45\textwidth]{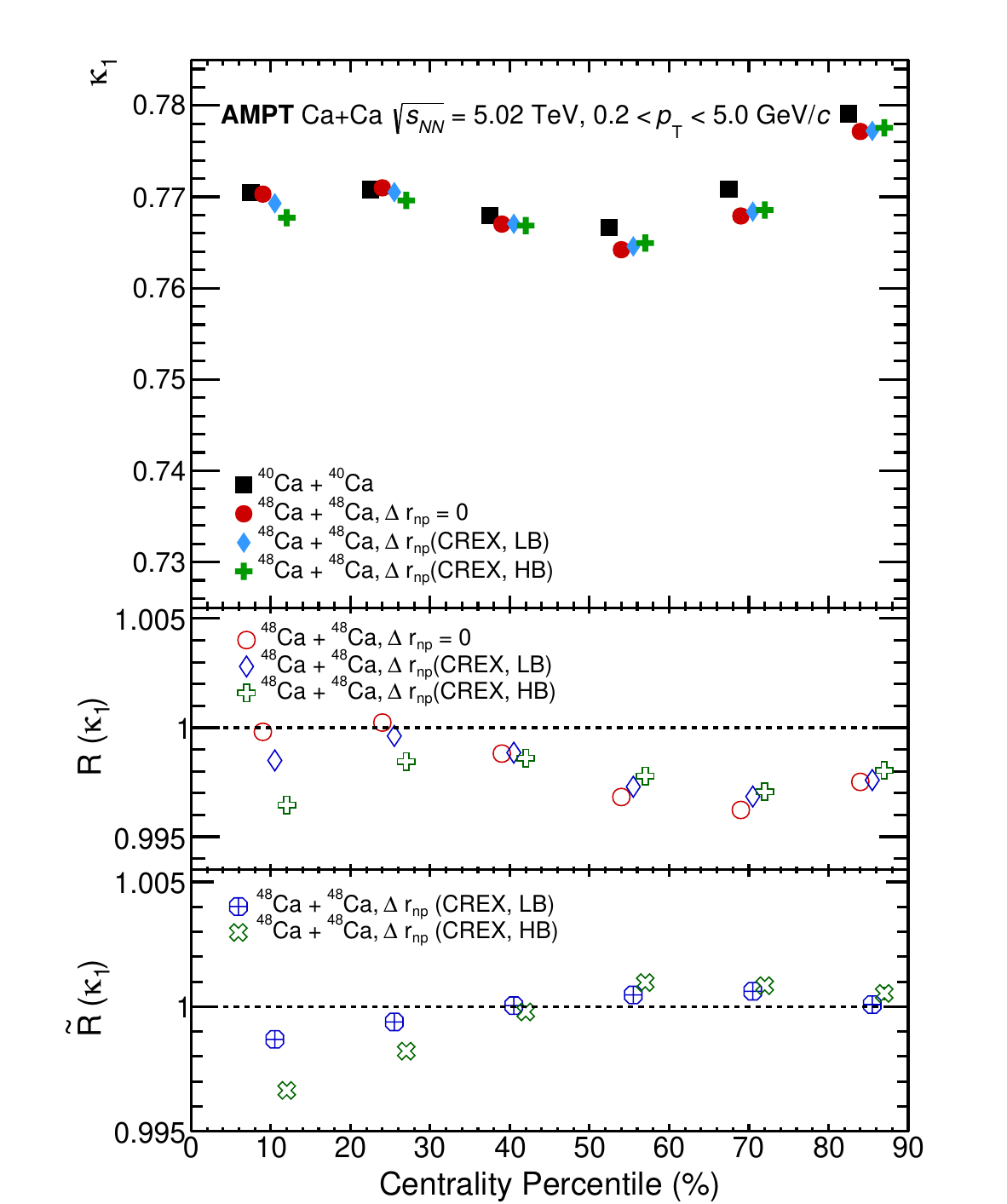}
\caption{\label{fig:k1} Centrality dependence of $\kappa_1$ in Ca+Ca collisions (top), the ratios of each $^{48}$Ca case over the $^{40}$Ca case (middle) and the ratios with the $^{48}$Ca+$^{48}$Ca with $\Delta r_{np}=0$ (bottom).}
\end{figure}
 
Figure \ref{fig:k1} (top) shows the centrality dependence of the first-order cumulant $\kappa_1$ in $^{48}$Ca+$^{48}$Ca and $^{40}$Ca+$^{40}$Ca collisions at $\sqrt{s_{_{\rm NN}}} = 5.02$ TeV from AMPT. A weak centrality dependence is observed for all $\kappa_1$ results, which are of similar magnitude to those obtained in collisions of other nuclei~\cite{Xu:2021uar,Nielsen:2023znu}. 
For the centrality range of 0--30\%, there seems to be a hierarchy of $\kappa_1$ from various neutron skin sizes. This trend becomes clearer when examining the $R(\kappa_1)$ ratios in the middle panel of Fig.~\ref{fig:k1}. It is expected that a larger neutron skin corresponds to a larger nucleus surface area, and gives a larger overlapping region of the two colliding nuclei. In central collisions, a larger overlapping region with the same number of nucleons gives a weaker pressure gradient, resulting in a smaller $\kappa_1$ result. Interestingly, this hierarchy might reverse when moving from central to peripheral collisions; nevertheless, the $\kappa_1$ result from the $^{40}$Ca case remains the largest for the presented centrality ranges. 
The results of the $R(\kappa_1)$ (middle) and $\tilde{R}(\kappa_1)$ (bottom) ratios show differences of less than $0.5\%$. In contrast, the current systematic uncertainties of the $\kappa_1$ measurements already exceed $0.5\%$~\cite{Giacalone:2023cet, ATLAS:2024jvf, ALICE:2010syw}. Even with future isotope runs of Ca nuclei at the LHC under identical experimental conditions, achieving a precision better than $0.5\%$ appears unlikely. Therefore, $\kappa_1$ is not an ideal observable for probing the neutron skin effect of $^{48}$Ca at the LHC.

\begin{figure}[t]
\includegraphics[width = 0.45\textwidth]{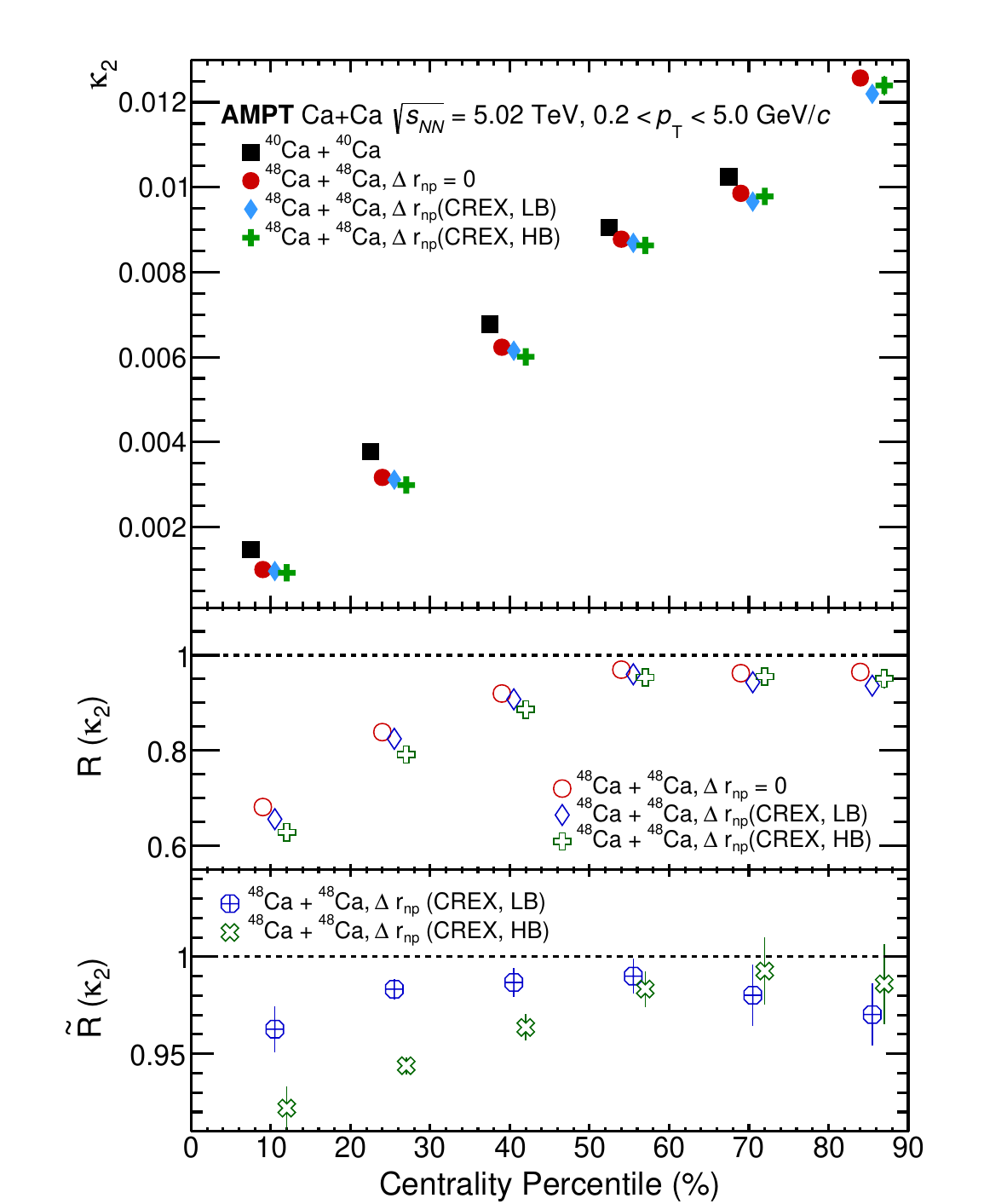} 
\caption{\label{fig:k2} Centrality dependence of $\kappa_2$ in Ca+Ca collisions (top), the ratios of each $^{48}$Ca case over the $^{40}$Ca case (middle) and the ratios between the $^{48}$Ca cases (bottom).}
\end{figure}

\begin{figure}[t]
\includegraphics[width = 0.45\textwidth]{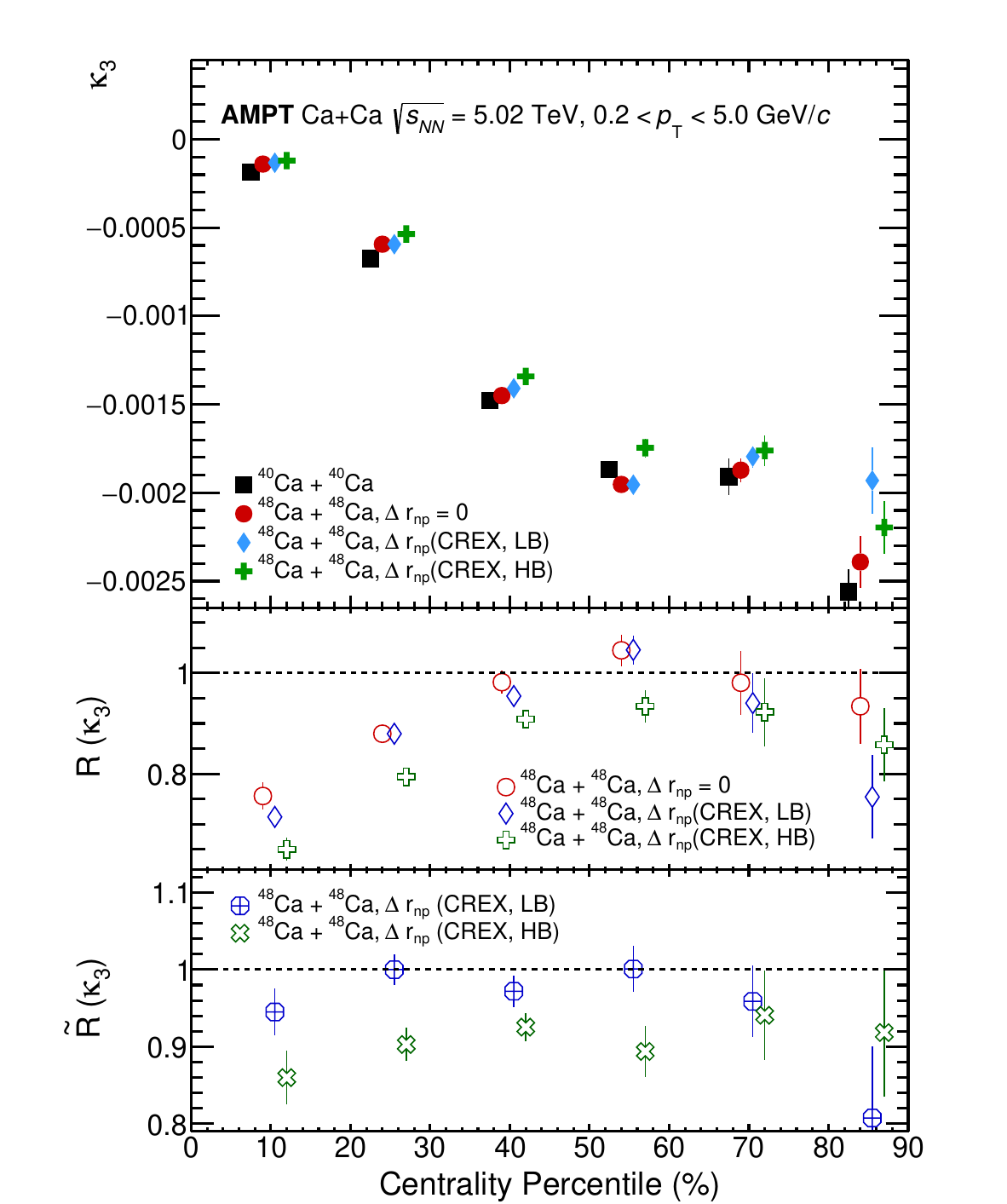} 
\caption{\label{fig:k3} Centrality dependence of $\kappa_3$ in Ca+Ca collisions (top), the ratios of each $^{48}$Ca case over the $^{40}$Ca case (middle) and the ratios with the $^{48}$Ca+$^{48}$Ca with $\Delta r_{np}=0$ (bottom).}
\end{figure}

Similarly, the centrality dependence of variance $\kappa_2$ and skewness $\kappa_3$ are studied using the AMPT model; the results are presented in Figs.~\ref{fig:k2} and \ref{fig:k3}, respectively. The top panel shows a positive result of $\kappa_2$, while the opposite is seen for $\kappa_3$. These trends are also observed in other AMPT studies of Xe+Xe collisions ~\cite{Nielsen:2023znu}, which differ from the experimental observations discussed above. 
Nevertheless, differences are observed in both $\kappa_2$ and $\kappa_3$ results, arising from variations in the neutron-skin settings for central and semi-central collisions. More specifically, $\kappa_2$ is inversely related to the neutron skin size, a trend that is better seen in the $R(\kappa_2)$ ratios shown in the middle panel of Fig.~\ref{fig:k2}. This behaviour can be understood as follows: for the same nucleon number, a smaller neutron-skin size produces a steeper pressure gradient, which in turn leads to stronger transverse momentum fluctuations, quantified by $\kappa_2$. A similar interpretation applies to $\kappa_3$.
In the central collisions, the $R(\kappa_2)$ and $R(\kappa_3)$ ratios deviate significantly from unity, more than 30\% and 20\%, respectively.  Among the results from different neutron-skin settings, the $R(\kappa_2)$ and $R(\kappa_3)$ ratios for the $\Delta r_{np}=0$ case (red circles) are particularly notable. For nuclei of the same size but with eight additional neutrons, the $^{48}$Ca+$^{48}$Ca collisions with $\Delta r_{np}=0$ exhibit significantly weaker mean transverse momentum fluctuations compared to those in $^{40}$Ca+$^{40}$Ca collisions, highlighting the role of initial nuclear density effects.

To better probe the neutron-skin effect, the $\tilde{R}$ ratios for $\kappa_2$ and $\kappa_3$ are shown in the bottom panels of Figs.~\ref{fig:k2} and \ref{fig:k3}. For the 0--30\% centrality range, the difference in $\kappa_2$ between $\Delta r_{np}(\text{CREX, LB})$ and $\Delta r_{np}=0$ is the smallest, about $3\%$, while the difference in $\kappa_2$ between $\Delta r_{np}(\text{CREX, HB})$ and $\Delta r_{np}(\text{CREX, LB})$ is slightly larger. At the same time, the $R(\kappa_2)$ ratios for $\Delta r_{np}=0$ and $\Delta r_{np}(\text{CREX, HB})$ show significant deviations from unity, reaching about $8\%$ in central collisions. Similar conclusions hold for the $\tilde{R}(\kappa_3)$ ratios. The observed differences across neutron-skin settings confirm that $\kappa_2$ and $\kappa_3$ are very promising observables for determining neutron-skin thickness. Notably, current systematic uncertainties in experiments are below 6$\%$ for $\kappa_2$ and below 3$\%$ for $\kappa_3$ in Pb+Pb collisions~\cite{ALICE:2023tej,ATLAS:2024jvf}. If approved and carried out under similar (or even identical) experimental conditions, future isotope runs of Ca at the LHC could further reduce the systematic uncertainties of $\kappa_2$ and $\kappa_3$, as well as those of ratio measurements, thereby opening new possibilities to resolve the current tensions between CREX and PREX results.

\begin{figure}[t]
\includegraphics[width = 0.45\textwidth]{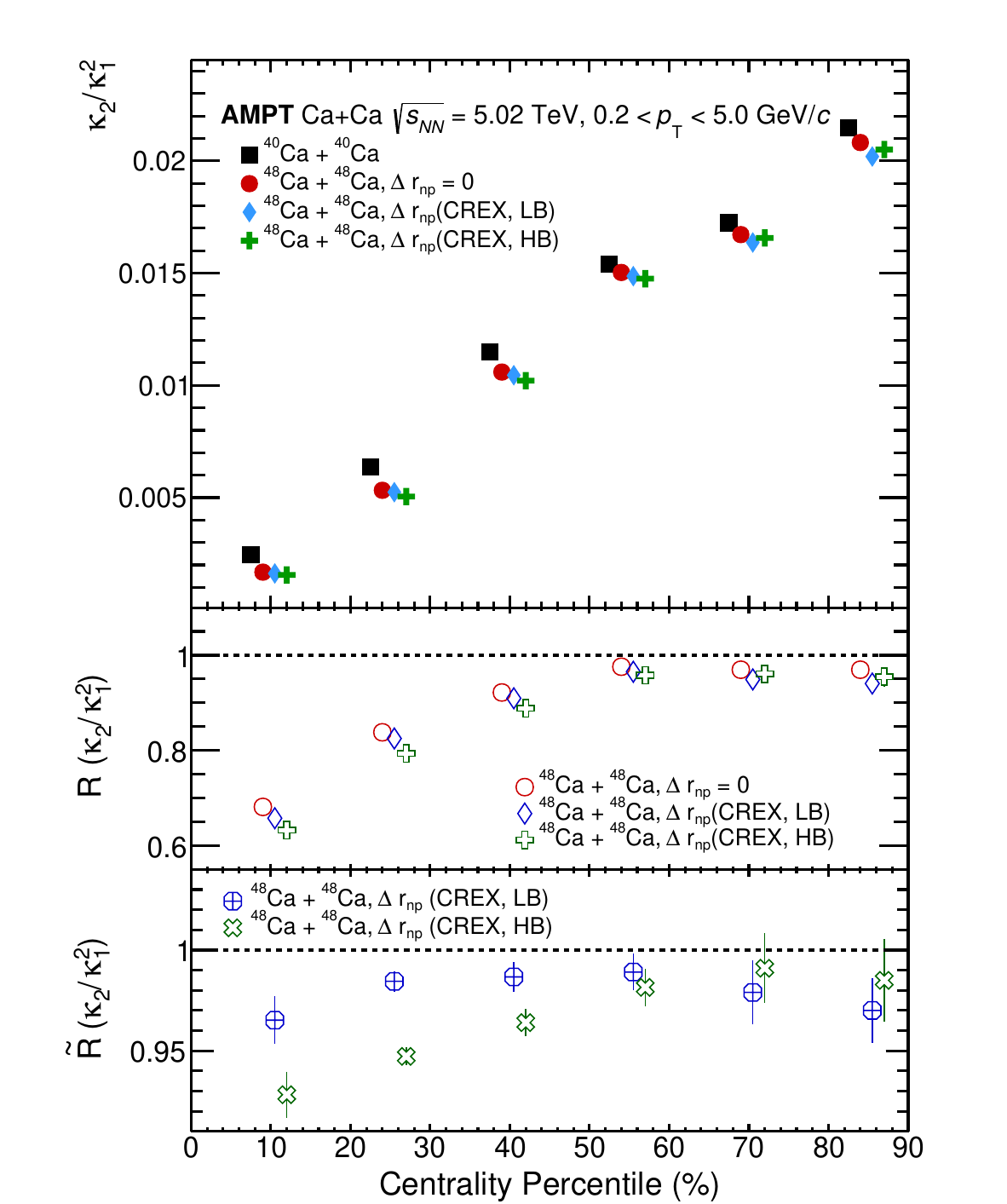}
\caption{\label{fig:k2_over_k1} Centrality dependence of $\kappa_2/\kappa^2_1$ in Ca+Ca collisions (top), the ratios of each $^{48}$Ca case over the $^{40}$Ca case (middle) and the ratios with the $^{48}$Ca+$^{48}$Ca with $\Delta r_{np}=0$ (bottom).}
\end{figure}

\begin{figure}[t]
\includegraphics[width = 0.45\textwidth]{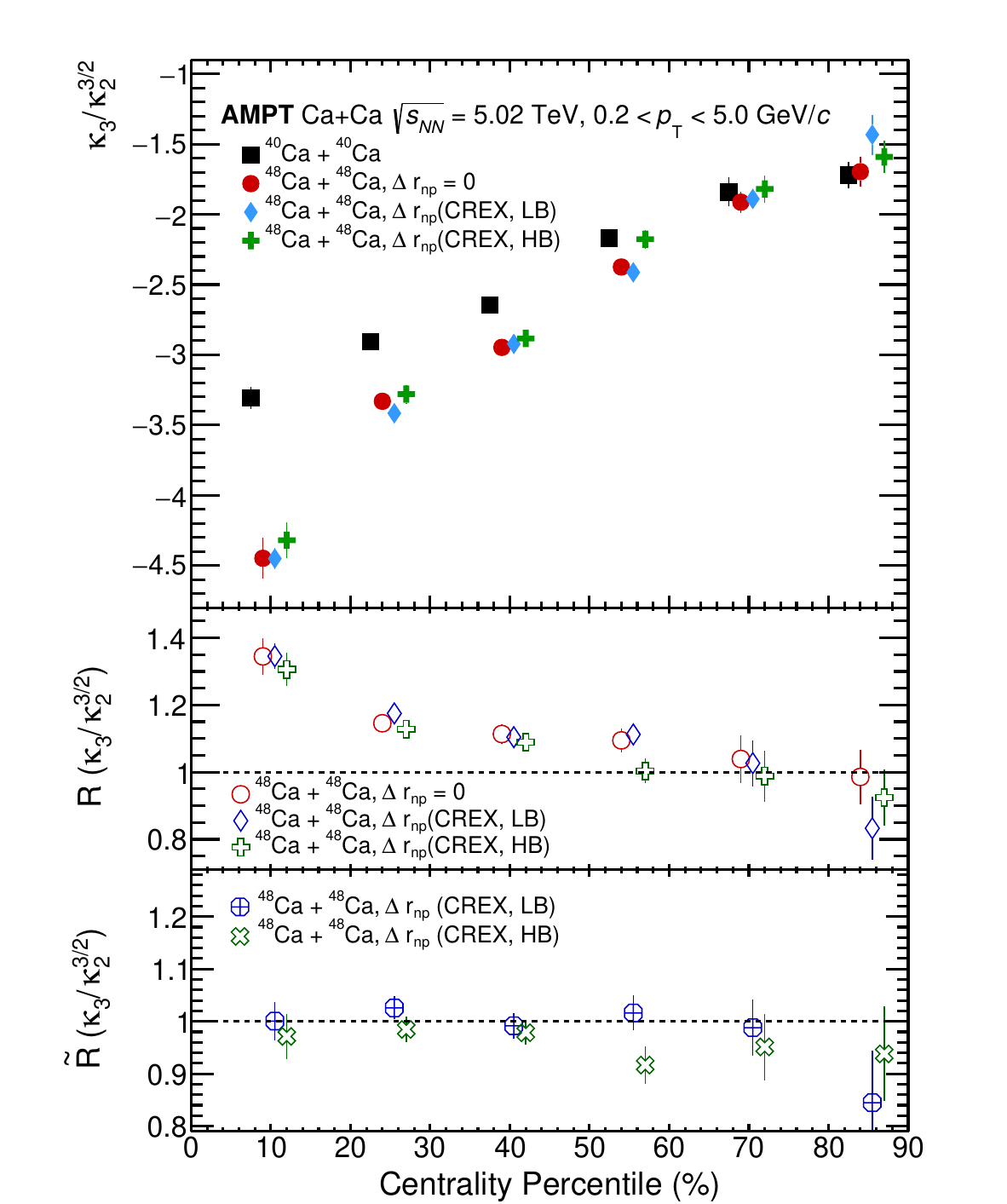}
\caption{\label{fig:k3_over_k2} Centrality dependence of $\kappa_3/\kappa^{3/2}_2$ in Ca+Ca collisions (top), the ratios of each $^{40}$Ca case over the $^{48}$Ca case (middle) and the ratios with the $^{48}$Ca+$^{48}$Ca with $\Delta r_{np}=0$ (bottom).}
\end{figure}

Despite the strong sensitivity of $\kappa_2$ and $\kappa_3$ to neutron-skin effects presented above, they are also highly sensitive to the complex dynamical evolution of ultra-relativistic nuclear collisions. This can introduce non-negligible uncertainties in the extracted neutron-skin thickness when comparing experimental measurements with theoretical model calculations. To mitigate such uncertainties, ratio observables have been proposed, as they largely cancel final-state effects~\cite{Giacalone:2020lbm} and provide improved access to the initial conditions and the neutron skin. 
The results of the ratio observable $\kappa_2/\kappa_1^{2}$ are presented in Fig.~\ref{fig:k2_over_k1}. The centrality dependence of $\kappa_2/\kappa_1^{2}$ is similar to that of $\kappa_2$, shown in Fig.~\ref{fig:k2}. Additionally, the differences between various neutron-skin settings can be seen in both $R(\kappa_2/\kappa_1^{2})$ and $\tilde{R}(\kappa_2/\kappa_1^{2})$ ratios. This similarity is expected, since the $\kappa_1$ results do not exhibit differences larger than $0.5\%$. After normalization by $\kappa_1^{2}$, the differences between neutron-skin settings remain significant, reaching about $7\%$ when comparing the results for $\Delta r_{np}=0$ with the case of $\Delta r_{np}(\text{CREX-HB})$. These differences are expected to be less sensitive to the system’s evolution and thus enable a more reliable extraction of the neutron-skin thickness in future studies.

The centrality dependence of $\kappa_3/\kappa_2^{3/2}$ is presented in Fig.~\ref{fig:k3_over_k2}. Unlike the similarity observed in the $\kappa_2/\kappa_1^{2}$ results, here the centrality dependence differs markedly from the $\kappa_3$ results shown in Fig.~\ref{fig:k3}. The magnitude of $\kappa_3/\kappa_2^{3/2}$ varies significantly between $^{40}$Ca and $^{48}$Ca. In the central collisions, the $R(\kappa_3/\kappa_2^{3/2})$ ratios differ from unity by about 30\%. Concerning the difference between $\Delta r_{np}=0$ in $^{48}$Ca+$^{48}$Ca and $^{40}$Ca+$^{40}$Ca, it highlights the effects of initial nuclear density, arising from nuclei of the same size but with different nucleon numbers. As collisions become more peripheral, these differences decrease and eventually vanish. In contrast to $\kappa_3$, the $\tilde{R}(\kappa_3/\kappa_2^{3/2})$ ratios are consistent with unity within sizeable uncertainties. This suggests that the $\kappa_3/\kappa_2^{3/2}$ ratio may not provide additional information on neutron-skin effects beyond what is already accessible through $\kappa_2$ and $\kappa_3$ observables.

Note that the current AMPT model (version 2.28), which is publicly available, has been observed to produce an inaccurate centrality dependence of the mean transverse momentum and its fluctuations in heavy-ion collisions~\cite{Nielsen:2023znu}. This issue has been addressed in an improved AMPT model~\cite{Zhao:2024feh}, which, unfortunately, is not publicly available. The present study on neutron skin effects via mean transverse momentum fluctuations should not be influenced by precise centrality dependence, but rather by relative comparisons between results from different neutron skin settings. Therefore, the conclusions are expected to remain robust. 

\section{Conclusions}

Utilising the ``imaging-by-smashing'' technique in proposed isotope runs of $^{40}\mathrm{Ca}+^{40}\mathrm{Ca}$ and $^{48}\mathrm{Ca}+^{48}\mathrm{Ca}$ collisions at the LHC offers a novel opportunity to investigate the neutron-skin effects of $^{48}\mathrm{Ca}$ through collective flow phenomena in AMPT simulations. Despite the strong sensitivity to the initial spatial nucleon distribution inside the nuclei, the results of elliptic flow from two-particle cumulant $v_2\{2\}$ have not yet shown a clear dependence on the neutron skin thickness $\Delta r_{np}$ based on the presented study. In contrast, the triangular and quadrangular flow results from the two-particle cumulant, $v_3\{2\}$ and $v_4\{2\}$, exhibit promising sensitivity to $\Delta r_{np}$. In addition to systematic investigations of anisotropic flow, the radial flow, characterised by event-by-event fluctuations in the mean transverse momentum, has been used to probe neutron skin effects. The second- and third-order cumulants of the mean transverse momentum fluctuation, $\kappa_2$ and $\kappa_3$, not only reveal a strong impact from changes in the initial density but also show significant variations in central collisions when $\Delta r_{np}$ is varied in the AMPT simulations. These variations, being considerably larger than current experimental precision, demonstrate that the neutron skin thickness leaves a unique imprint on the final-state $\kappa_2$ and $\kappa_3$ observables. Since the proposed isotope runs under identical experimental conditions would largely mitigate both theoretical and experimental uncertainties, future studies of $\kappa_2$ and $\kappa_3$, as well as the normalised ratio $\kappa_2/\kappa_1^2$, will place tight constraints on the extracted neutron skin thickness. 

The study presented in this work focuses on investigating the neutron skin thickness through variations in the $R_{n}$ parameter alone. A complementary study of the diffuseness parameter $a_{0}$ would also be very useful. This has been partially explored via elliptic flow study in the isobar runs of $^{96}$Ru+$^{96}$Ru and $^{96}$Zr+$^{96}$Zr at RHIC~\cite{Jia:2022qgl} and also $^{129}$Xe+$^{129}$Xe at the LHC~\cite{ALICE:2024nqd,Mantysaari:2024uwn}, where the flow results show significant sensitivity to the changes in $a_{0}$. In addition, the study based on density functional theory seems to favour the halo-type instead of skin-type effect of $^{76}$Zr~\cite{Li:2018oec}. To obtain a reliable and precise determination of the neutron-skin parameters $a_0$ and $R_n$ of $^{48}$Ca, it will be essential to perform global Bayesian analyses~\cite{Paquet:2023rfd, Giacalone:2023cet} that use future measurements of anisotropic flow and mean transverse-momentum fluctuations as key inputs. The maximum physics impact will be achieved if $^{48}$Ca+$^{48}$Ca and $^{40}$Ca+$^{40}$Ca collisions can be taken under identical experimental conditions~\cite{AlemanyFernandez:2025ixd}.
Ultimately, the application of the imaging-by-smashing technique to the proposed isotope runs on Ca nuclei may help resolve the current tension over the neutron skin and the symmetry energy. Meanwhile, its connection to low-energy nuclear physics and astrophysics will significantly broaden the scientific scope of the ultra-relativistic nuclear collision program at the CERN Large Hadron Collider.

\section*{Acknowledgements}

We thank Xinli Zhao for her valuable help on the AMPT simulations, and thank Giuliano Giacalone and Haojie Xu for useful feedback on the manuscript. We would also like to thank the organisers and participants of the workshop `Light Ion Collisions at the LHC (11-15 November 2024, CERN)', where the works were discussed in the early stages. The authors are supported by the European Union (ERC, InitialConditions, 101077147), the VILLUM FONDEN (grant number 00025462), and the Independent Research Fund Denmark (DFF-Sapere Aude grant, 2064-00052).

\printbibliography

\end{document}